\documentclass[11pt]{article}
\usepackage{amsfonts}
\usepackage{amssymb}
\usepackage{epic}
\usepackage[english,francais,german]{babel}

\begin{document}
\selectlanguage{english}
\thispagestyle{empty}

\begin{center}
{\Large\bf On Wolfgang Pauli's most important contributions to physics\footnote{
Opening talk at the symposium: WOLFGANG PAULI AND MODERN PHYSICS, 
in honor of the 100th anniversary of Wolfgang Pauli's birthday, ETH (Zurich), Mai 4-6 2000.}}

\vspace{.5cm}

{{\bf Norbert Straumann}\\
 Institute for Theoretical Physics, University of Zurich,\\
 Winterthurerstrasse 190, 8057 Z\"urich, Switzerland} 

\vspace{.5cm}

\begin{abstract}
After some brief biographical notes, Wolfgang
Pauli's main contributions to general relativity,
quantum theory, quantum field theory, and
elementary particle physics are reviewed. A
detailed description is given how Pauli
descovered -- before the advent of the new quantum
mechanics -- his exclusion principle.
\end{abstract}

\end{center}

\vspace{2.5cm}
\selectlanguage{german}

\noindent
Meine Damen und Herren,

\vspace{.5cm}

\noindent
Ich f\"uhle mich sehr geehrt, diesen Vortrag halten zu
d\"urfen. Gleichzeitig habe ich ein etwas beklemmendes Gef\"uhl, denn
meine Aufgabe ist nicht leicht.

Am Samstag werden wir manches \"uber Paulis Pers\"onlichkeit und seine 
weitgespannten Interessen h\"oren. In meinem Vortrag werde ich
hingegen fast ausschliesslich \"uber Paulis Wissenschaft$^{[1]}$ reden.

Auch bei \"ausserster Konzentration auf seine wesentlichsten
Beitr\"age ist eine Stunde nat\"urlich zu kurz. Deshalb muss ich die
Gewichte sehr ungleichm\"assig verteilen. 

\medskip

Ich sollte zumindest auf folgende Themen eingehen:

\begin{enumerate}
\item[1)] Zu Beginn seiner Karriere hat sich Pauli in sehr jungen Jahren
  mit seinen Beitr\"agen zur Allgemeinen Relativit\"atstheorie,
  insbesondere zur Weylschen Erweiterung, sehr schnell einen Namen als 
  ganz aussergew\"ohnlicher theoretischer Physiker gemacht. 
\item[2)] Paulis wichtigster Beitrag zur Physik ist zweifellos sein
  Ausschliessungsprinzip. Der wenig bekannten Entdeckungsgeschichte
  werde ich einen wesentlichen Teil meiner Zeit widmen. Die
  Schwierigkeit dabei ist, dass man sich in die Situation {\it vor}
  der Quantenmechanik (in unserem heutigen Sinne) zur\"uckversetzen
  muss. Nach genauem Studium von Paulis Argumentationskette bin ich
  zutiefst davon beeindruckt, wie Pauli --- auf der Basis der br\"uchigen 
  Bohr-Sommerfeld-Theorie und dem damaligen spektroskopischen Wissen
  --- der Natur sein Ausschliessungsprinzip ablauschen konnte. Die
  Genialit\"at des jungen Mannes zeigt sich, so denke ich, nirgends
  deutlicher. 
\item[3)] Die weiteren wichtigsten Beitr\"age zur Quantenmechanik, zur 
  Quantenfeldtheorie und zur Elementarteilchenphysik werde ich relativ 
  knapp abhandeln m\"ussen. Dies ist nicht so schlimm, da man w\"ahrend 
  des Studiums mit diesen Themen konfrontiert wird und Paulis
  Einsichten zum selbstverst\"andlichen Besitz der Physik geworden
  sind.

Auf den Zusammenhang von {\it Spin und Statistik} muss ich aber
unbedingt etwas eingehen, da mit diesem --- zur grossen Befriedigung
von Pauli --- das Ausschlussprinzip eine tiefere Begr\"undung fand.

Ferner m\"ochte ich etwas zur {\it Neutrinohypothese} sagen, vor allem 
weil die damit verbundenen anf\"anglichen Schwierigkeiten meist
ungen\"ugend hervorgehoben werden. Es ging eben nicht nur um die
Energieerhaltung, sondern u.a. auch um Spin und Statistik der Atomkerne.
\end{enumerate}

\centerline{\Large *~~~*~~~*~~~*~~~*}

\vspace{.5cm}

\section*{Werdegang}

In einem Alter, in welchem heute die meisten ihre Dissertation
aufschreiben, war Pauli bereits eine der wichtigsten
Autorit\"aten und treibenden Kr\"afte in der Welt der theoretischen
Physik. ``Was, so jung und {\it schon} unbekannt'' soll Pauli einmal
zu einem jungen Mann gesagt haben. 

Paulis Ausstrahlung auf die Physiker seiner Zeit hat Paul Ehrenfest
anl\"asslich der Verleihung der Lorentzmedaille an Pauli  
im Jahre 1931 so zusammengefasst$^{[2]}$:
\begin{quote}
``Aber vielleicht noch von viel gr\"osserem Gewicht als seine
Publikationen sind die unz\"ahligen, unverfolgbaren Beitr\"age, die er 
zur Entwicklung der neueren Physik durch m\"undliche Diskussionen oder 
Briefe geliefert hat. Die enorme Sch\"arfe seiner Kritik, seine
ausserordentliche Klarheit und vor allem die r\"ucksichtslose
Ehrlichkeit, mit der er stets den Nachdruck auf die ungel\"osten
Schwierigkeiten legt, bewirkt, dass er als unsch\"atzbare Triebkraft
innerhalb der neueren theoretischen Forschung gelten muss.''
\end{quote}
\noindent
Einen lebendigen Eindruck von Paulis Einfluss auf die Physiker seiner
Zeit und seine unerh\"ort kritische F\"ahigkeit gibt der nie
abreissende Briefstrom$^{[3]}$, bei dessen Lekt\"ure man immer wieder den
Atem anh\"alt. Paulis Kritik war gesucht, gef\"urchtet und oft
beissend, wie die folgende etwas erschreckende Kostprobe aus einem
Brief an Einstein$^{[4]}$ aus dem Jahre 1929 zeigt, der gleichzeitig auch den
stattgefundenen Generationswechsel \"uberdeutlich macht (als
Hintergrund f\"ur diesen  Brief muss man wissen, dass Einstein bereits 
in den 20er Jahren mehrere Versuche machte, Gravitation und
Elektrodynamik in einer einheitlichen klassischen Feldtheorie zu
verschmelzen, wovon Pauli wenig hielt):
\begin{quote}
``Es bleibt [den Kritik \"ubenden Physikern] nur \"ubrig, Ihnen dazu
zu gratulieren (oder soll ich lieber sagen: zu kondolieren?), dass Sie 
zu den reinen Mathematikern \"ubergegangen sind. Ich bin auch nicht so 
naiv als dass ich glauben w\"urde, Sie w\"urden auf Grund irgend einer 
Kritik durch Andere Ihre Meinung \"andern. Aber ich w\"urde jede Wette 
mit Ihnen eingehen, dass Sie sp\"atestens nach einem Jahr den ganzen
Fernparallelismus aufgegeben haben werden, so wie Sie fr\"uher die
Affintheorie aufgegeben haben. Und ich will Sie nicht durch
Fortsetzung dieses Briefes noch weiter zum Widerspruch reizen, um das
Herannahen dieses nat\"urlichen Endes der Fernparallelismustheorie
nicht zu verz\"ogern.''
\end{quote}
\noindent
Pauli war, was man ein Wunderkind nennt. Als reifer Mann meinte er
dazu: ``Ja, das Wunderkind --- das Wunder vergeht und das Kind bleibt
\dots~.'' Es war wohl naheliegend, dass Pauli nach dem Abitur noch im
Jahre 1918 nach M\"unchen zog, wo Sommerfeld `eine Pflanzst\"atte der 
theoretischen Physik' geschaffen hatte und die besten Talente
anzog. Sommerfeld ist vor allem als begnadeter Lehrer in die
Geschichte der Physik eingegangen. Seine auch heute noch sehr
lesenswerten Lehrb\"ucher geben uns eine Ahnung von der Begeisterung,
mit der er seine Vorlesungen hielt und mit welchem Enthusiasmus er die 
jugendlichen H\"orer mitriss. Zu den Massst\"aben seines Unterrichts
sagte er einmal$^{[5]}$: ``Was wir betonen m\"ussen, ist die ideale Seite
der mathematischen und naturwissenschaftlichen Studien, ihre
Sch\"onheit und innere Wahrhaftigkeit; ihre Kraft, phrasenloses Denken 
und r\"ucksichtsloses Schliessen im Sch\"uler zu entwickeln.''

Sommerfeld erkannte sofort die Begabung von Pauli und betraute ihn
schon in den Anfangssemestern mit dem f\"ur die Enzyklop\"adie der
mathematischen Wissenschaften bestimmten Kapitel \"uber
Relativit\"atstheorie. 
Zu diesem Zeitpunkt geh\"orte Pauli bereits zu den besten Kennern 
der Allgemeinen Relativit\"atstheorie. Im Jahre 1919 sind von ihm drei 
Arbeiten$^{[6]}$ auf diesem Gebiet erschienen, wovon zwei der kurz
zuvor entstandenen Weylschen Theorie von Gravitation und
Elektrizit\"at gewidmet sind.

Die Weylsche Theorie war --- mathematisch gesehen --- ein bestechender 
Versuch, die beiden damals einzig bekannten Wechselwirkungen in
geometrischer Weise einheitlich zu beschreiben. Weyl verallgemeinerte
dabei die Riemannsche Geometrie auf sehr nat\"urliche Weise, wodurch
nicht nur das Gravitationsfeld sondern auch das elektromagnetische
eine geometrische Bedeutung erlangten$^{[7]}$.

Einstein \"ausserte mit Recht sofort physikalische Bedenken, aber
Pauli war nach Weyl der erste, der die Theorie n\"aher untersuchte und 
beobachtbare Konsequenzen ausarbeitete. So berechnete er die
Periheldrehung des Merkur und die Lichtablenkung in der Weylschen
Theorie f\"ur eine damals von Weyl bevorzugte Wirkung\footnote{
Da die zugeh\"origen Feldgleichungen h\"oherer Ordnung sind,
tritt im zentralsymmetrischen Problem neben der Masse noch
eine zweite willk\"urliche Integrationskonstante auf.
Pauli zeigt, dass nur f\"ur eine spezielle Wahl dieser
Integrationskonstante die Einsteinschen Resultate f\"ur
die Periheldrehung und die Lichtablenkung resultieren.}.

In diesen ersten Arbeiten von Pauli sieht man bereits den ausgereiften
Meister. Jedermann wunderte sich, wie dies so fr\"uh m\"oglich
war$^{[8]}$.

Pauli befand sich erst im dritten Semester als er im Sommer 1920 auf
Veranlassung von Sommerfeld mit der Arbeit an dem f\"ur die
Enzyklop\"adie der mathematischen Wissenschaften bestimmten Kapitel
\"uber Relativit\"atstheorie begann. In weniger als einem Jahr
bew\"altigte er --- neben seinem Studium --- die anspruchsvolle
Aufgabe und lieferte Anfang 1921 das Manuskript von 237 Buchseiten mit 
gegen 400 verarbeiteten Literaturzitaten ab. Damit etablierte sich
Pauli als ein Wissenschaftler von seltener Tiefe und alles
\"uberragenden synthetischen und kritischen F\"ahigkeiten. In seiner
Besprechung des Enzyklop\"adieartikels$^{[9]}$ schrieb Einstein$^{[10]}$:
\begin{quote}
``Wer dieses reife und gross angelegte Werk studiert, m\"ochte nicht
glauben, dass der Verfasser ein Mann von einundzwanzig Jahren ist. Man 
weiss nicht, was man am meisten bewundern soll, das psychologische
Verst\"andnis f\"ur die Ideenentwicklung, die Sicherheit der
mathematischen Deduktion, den tiefen physikalischen Blick, das
Verm\"ogen \"ubersichtlicher systematischer Darstellung, die
Literaturkenntnis, die sachliche Vollst\"andigkeit, die Sicherheit der 
Kritik.''
\end{quote}
\noindent
Im selben Jahr 1921, als Paulis Artikel erschien, promovierte er
bereits nach sechs Semestern an der Universit\"at M\"unchen mit einer
Arbeit \"uber das Wasserstoff-Molek\"ulion$^{[11]}$ im Rahmen der
alten Quantentheorie von Bohr und Sommerfeld. Die Grenzen dieser
Theorie zeichneten sich bereits bei diesem Problem ab. Vollends klar
wurde dies allerdings erst nachdem Heisenberg auch das Heliumatom
untersucht hatte. 

Nach seiner Promotion trat Pauli im Wintersemester 1921/22 eine
Assistentenstelle bei Max Born in G\"ottingen an. In gemeinsamen
Untersuchungen \"ubertrugen die beiden die astronomische
St\"orungstheorie auf die Atomphysik. Schon am 29.~November 1921
schrieb Born an Einstein$^{[12]}$: ``Der kleine Pauli ist sehr
anregend; einen so guten Assistenten werde ich nie mehr kriegen.''
Jahrzehnte sp\"ater \"ausserte sich Born \"uber Pauli
folgendermassen$^{[13]}$: ``Denn ich wusste seit der Zeit, da er mein
Assistent in G\"ottingen war, dass er ein Genie war, nur vergleichbar
mit Einstein selbst, ja dass er rein wissenschaftlich vielleicht noch
gr\"osser war als Einstein, wenn auch ein ganz anderer Menschentyp,
der in meinen Augen Einsteins Gr\"osse nicht erreichte.''

\vspace{.5cm}

\section*{Entdeckungsgeschichte des Ausschliessungsprinzips}

Einen lebendigen Eindruck von Paulis Entdeckung des
Ausschliessungsprinzips$^{[14]}$ --- seinem wichtigsten Beitrag zur
Physik --- gewinnt man wieder aus dem ersten Band des
Briefwechsels$^{[15]}$. Das Pauliprinzip lag zum Zeitpunkt seiner
allgemeinen Formulierung gegen Ende 1924 keineswegs in der Luft, stand 
man doch damals vor zwei grunds\"atzlichen Schwierigkeiten: Einen
allgemeinen \"Ubersetzungsschl\"ussel eines mechanischen Modells in
eine koh\"arente Quantentheorie gab es noch nicht und der
Spin-Freiheitsgrad war unbekannt. 

Die Entdeckungsgeschichte des Ausschliessungsprinzips beginnt im
Herbst 1922 in Kopenhagen, wo sich Pauli am Bohrschen Institut mit
einer Erkl\"arung des anomalen Zeeman-Effektes abm\"uhte. Land\'e
hatte aus den Spektraldaten bereits die Aufspaltung der Spektralterme
in einem schwachen Magnetfeld abgeleitet und f\"ur die Beschreibung
der Dublettstruktur der Alkalimetalle halbzahlige magnetische
Quantenzahlen eingef\"uhrt. Pauli gelang es in einem ersten Schritt,
Land\'es Termanalyse f\"ur das Paschen-Back-Gebiet zu
verallgemeinern. Er fand dabei das ``allgemeine formale Gesetz,
welches gestattet, die Aufspaltungsfaktoren $g$ im Falle kleiner
Felder aus den Energiewerten bei grossen Feldern
abzuleiten''$^{[16]}$. (Darauf werde ich gleich n\"aher eingehen.)

Diese fr\"uhe Arbeit war --- wie Pauli in seinem Nobelvortrag$^{[17]}$ 
betont --- f\"ur die Entdeckung des Ausschliessungsprinzips
wesentlich. Zum Zeitpunkt ihrer Abfassung war er aber dar\"uber sehr
ungl\"ucklich, wie z.~B. aus einem Brief an Sommerfeld
hervorgeht$^{[18]}$:
\begin{quote}
``Ich habe mich sehr lange mit dem anomalen Zeemaneffekt geplagt,
wobei ich oft auf Irrwege geriet und eine Unzahl von Annahmen pr\"ufte 
und dann wieder verwarf. Aber es wollte und wollte nicht stimmen! Dies 
ist mir bis jetzt einmal gr\"undlich schief gegangen! Ein Zeit lang
war ich ganz verzweifelt \dots ich habe das Ganze mit einer Tr\"ane im 
Augenwinkel geschrieben und habe davon wenig Freude.''
\end{quote}
\noindent
Pauli hat selber erz\"ahlt$^{[19]}$ wie ihm auf einer ziellosen
Wanderung durch die Strassen Kopenhagens Harald Bohr begegnete, der
freundlich zu ihm sagte: ``Sie sehen so ungl\"ucklich aus'', worauf er 
schroff antwortete: ``Wie kann man gl\"ucklich aussehen, wenn man an
den anomalen Zeemaneffekt denkt.''

\vspace{.5cm}

\noindent
{\bf 1.~Schritt: Zeemaneffekt in starken Feldern und Paulis
  Summenregel}$^{[16]}$

\medskip

\noindent
Ich m\"ochte Ihnen diesen ersten Schritt auf dem Weg zum
Ausschliessungsprinzip im Detail vorf\"uhren, um Ihnen zu zeigen, wie
Pauli auf unsicherem Grund zu einem richtigen --- und wie sich zeigen
sollte --- wichtigen Resultat gelangte. Dabei werde ich heutige
Bezeichnungen verwenden, um unn\"otige Schwierigkeiten zu vermeiden. 

Wie immer beschreibt Pauli zun\"achst in konziser Weise, was er an
bereits Vorhandenem zugrunde legt. Dies muss ich ebenfalls festhalten:

\begin{itemize}
\item Die Beziehung zwischen Energieniveaus und Spektrum wird durch die 
  {\it Bohrsche Regel}
\[
E_2-E_1=h\,\nu
\]
bestimmt. (Diese kannte man seit Bohrs Pionierarbeiten.)
\item Den Spektraltermen hatte man bereits verschiedene {\it
    Quantenzahlen} zugeordnet. Es sind dies\footnote
{In eckigen Klammern sind die historischen Bezeichnungen gegeben.}:
\begin{description}
\item[$\vartriangleright$] $L
  [=k-1]~,~~L=0,1,2,3,\ldots~~~(S,P,D,F,\ldots)$\\
(unsere heutige Bahndrehimpuls-Quantenzahl);
\item[$\vartriangleright$] $S [=i-1]$: ~jeder Term geh\"ort zu einem 
  Singlett oder Multiplettsystem mit {\it maximaler}
  Multiplizit\"at~ $2 S + 1~ (S=0, \frac 12, 1, \ldots )$ (unsere
  heutige Spinquantenzahl);
\item[$\vartriangleright$] die verschiedenen Terme eines Multipletts
  (mit denselben  
  $L$ und $S$) werden durch die Quantenzahl $J [=j]$ unterschieden;
  deren Werte sind:\\

$\phantom{mmmmmm}J=L+S,~L+S-1,\ldots~~ L-S$~~ f\"ur~~$L \geq S$~,\\
$\phantom{mmmmmm}J=S+L,~S+L-1,\ldots~~ S-L$~~ f\"ur~~$L < S$~.\\

($J$ ist nat\"urlich unsere heutige totale Drehimpulsquantenzahl.)
\end{description}
\item Ferner kannte man die folgenden {\it Auswahlregeln}, die in den
  meisten F\"allen g\"ultig sind:
\begin{eqnarray}
&&L\;\longrightarrow\; L\pm 1~,\nonumber \\ 
&&S\;\longrightarrow\;S~,\nonumber\\
&&J\;\longrightarrow\;J+1, J, J-1~~~(0~\longrightarrow ~0{\rm
  ~verboten}).\nonumber 
\end{eqnarray}
\item Bei gegebener Atomzahl $Z$ ($Z-p$ f\"ur $p$-fach ionisierte Atome)
  gilt:
\begin{eqnarray}
&&Z~~{\rm gerade}~~~~\longrightarrow~S, J\,:~~{\rm ganz}~,\nonumber\\
&&Z~~{\rm ungerade}~\longrightarrow~S, J\,:~{\rm
  halbganz}~.~~~~~~~~~~~~~\nonumber 
\end{eqnarray}
\item {\it Aufspaltung im Magnetfeld:}
\begin{description}
\item[$\vartriangleright$] Jeder Term spaltet in $2J+1$ Niveaus auf,
  welche durch die Quantenzahl $M=J, J-1, \ldots, -J$ unterschieden
  werden. 
\item[$\vartriangleright$] {\it Land\'e:} F\"ur schwache Felder ist
  die Aufspaltung ~ $\triangle E_M = M \,\cdot\,g (\mu_0 B)$~
  (\"aquidistante Reihe); dabei ist ~$\mu_0 = e\hbar/2mc$~ das Bohrsche 
  Magneton (von Pauli 1920 eingef\"uhrt) und $g$ ist der Land\'esche
  $g$-Faktor: 
\[
g\;=\;\frac 3 2 \;+\;\frac{S(S+1) - L(L+1)}{2J (J+1)}~;
\]
\item[$\vartriangleright$] {\it Auswahlregeln} f\"ur
  Zeeman-\"Uberg\"ange:
\begin{eqnarray}
&&M\;\longrightarrow\;M\pm 1~~(\sigma-{\rm Komponente}),\nonumber\\
&&M\;\longrightarrow\;M~~~~~~~ (\pi-{\rm Komponente}).~~~~~~~\nonumber
\end{eqnarray}
{\it Bemerkung:} Falls $g$ und $g'$ im Anfangs- und Endzustand gleich
sind gilt:

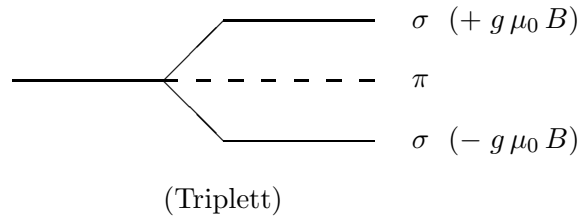
\setlength{\unitlength}{1mm}
\begin{picture}(150,40)(0,-20)
  \thinlines
  \put(20,0){\line(1,0){20}}
  \put(40,0){\line(1,1){8}}
  \put(40,0){\line(1,-1){8}}
  \put(48,8){\line(1,0){20}}
  \put(48,-8){\line(1,0){20}}
  \dashline{2}(40,0)(68,0)
  \put(73,8){\makebox(0,0)[s]{$\sigma~~~(+\;g\,\mu_0\,B)$}}
  \put(73,0){\makebox(0,0)[s]{$\pi$}}
  \put(73,-8){\makebox(0,0)[s]{$\sigma~~~(-\;g\,\mu_0\,B)$}}
  \put(40,-16){\makebox(0,0)[s]{(Triplett)}}
\end{picture}

\end{description}
\end{itemize}


\noindent 
Pauli akzeptiert diese {\it empirischen Regeln} und untersucht das
experimentelle Material f\"ur {\it starke} Felder. In Tabellenform
pr\"asentiert er die Aufspaltungen als Vielfache von $\mu_0B$ und
beschreibt das Ergebnis folgendermassen:

F\"uhrt man zwei Zahlen $M_L, M_S, [= m_1,\mu]$ ein, deren Summe
gleich $M$ ist, 
\[
M\;=\;M_L+M_S~,~~~~~~~~~~~~~~~~
\]
und die die folgenden Werte annehmen
\[
M_L\;=\;L, L-1, \ldots, -L~,~~~~~~~~~~~~~~~~~~~~~~~~~~~~~~
\]
\[
M_S\;=\;\left\{
\matrix{ &\!\!\!\!\!\pm\;\frac 12 &\!\!\!\!\!\!\!\!\!{\hbox{f\"ur~Dubletts~(Alkalien)}}\cr
         &0,\;\pm\; 1   &{\hbox{f\"ur~Tripletts~(Edalkalien)}~,}}\right. 
\]
so gilt die einfache Formel:
\[
\triangle E/\mu_0 B\;=\;M_L+2M_S\;=\;M+M_S~.
\]
Dies {\it verallgemeinert} Pauli sofort auf {\it h\"ohere Multipletts}
durch die Annahme, dass dieselben Formeln gelten, aber $M_S$ die Werte
\[
S, S-1, \ldots , - S~~~~~~~~~~~~~~~~
\]
durchl\"auft. Diese Verallgemeinerung war damals noch nicht
experimentell gepr\"uft.

Die Auswahlregel f\"ur $M_S$ ist: ~$M_S \to M_S$; deshalb ist der
Zeemaneffekt bei starken Feldern {\it normal}.

Nun postuliert Pauli eine bemerkenswerte formale Regel, die es gestattet
die Land\'eschen $g$-Faktoren im Falle schwacher Felder aus den
Energiewerten bei grossen Feldern abzuleiten. Paulis `Summensatz'
lautet:
\begin{quote}
``Die Summe der Energiewerte eines Multiplets, die zu gegebenen Werten 
von $M$ und $L$ geh\"oren, bleibt w\"ahrend des ganzen \"Ubergangs von 
schwachen zu starken Feldern eine {\it lineare} Funktion der
Feldst\"arke $B$.''
\end{quote}
(Es sei gleich bemerkt, dass dieser Summensatz in der QM richtig
ist\renewcommand{\thefootnote}{\fnsymbol{footnote}}\footnote[1]{Die
  Summe in Paulis Regel ist gleich der Spur von $\langle H_B\rangle$,
  $H_B=\mu_0B (J_3+S_3)$, wenn $\langle H_B\rangle$ die St\"ormatrix
  f\"ur festes $M$ bezeichnet. Diese Spur ist offensichtlich linear in 
  $B$.}\renewcommand{\thefootnote}{\arabic{footnote}} 
 .)
\medskip

\noindent
{\bf Spezialf\"alle}:
\begin{enumerate}
\item[1)] $M=J=L+S$. Daf\"ur gibt es nur {\it einen} Zustand, also ist 
  dessen Energie linear in $B$.
\item[2)] $M=2S+1$. Daf\"ur haben wir die {\it maximale} Zahl von
  $2S+1$ Zust\"anden; deren arithmetisches Mittel bei starken Feldern
  ist gleich $\mu_0B$. Aus dem Summensatz schliessen wir, dass {\it
    das arithmetische Mittel der $g$-Faktoren $=1$} ist. 
\end{enumerate}

\noindent
Pauli zeigt nun --- und darauf legt er sein Hauptgewicht ---, dass mit
seiner formalen Regel tats\"achlich alle $g$-Faktoren eindeutig aus den
Energiewerten bei {\it starken} Feldern berechnet werden
k\"onnen. Dazu wendet er den `Summensatz' der Reihe nach auf die
verschiedenen Werte von $M$ bei gegebenem $L$ an. Dies will ich nicht
ausf\"uhren.

Charakteristisch f\"ur Pauli ist der Schlussabschnitt$^{[16]}$:
\begin{quote}
``Eine befriedigende modellm\"assige Deutung der dargelegten
Gesetzm\"assigkeiten, insbesondere der in diesem Paragraphen
besprochenen formalen Regel ist uns nicht gelungen. Wie schon in der
Einleitung erw\"ahnt, d\"urfte eine solche Deutung auf Grund der
bisher bekannten Prinzipien der Quantentheorie kaum m\"oglich
sein. Einerseits zeigt das Versagen des Larmorschen Theorems, dass die 
Beziehung zwischen dem mechanischen und dem magnetischen Moment eines
Atoms nicht von so einfacher Art ist wie es die klassische Theorie
fordert, indem das Biot-Savartsche Gesetz verlassen oder der
mechanische Begriff des Impulsmomentes modifiziert werden
muss. Andererseits bedeutet das Auftreten von halbzahligen Werten von
$m$ und $j$ bereits eine grunds\"atzliche Durchbrechung des Rahmens
der Quantentheorie der mehrfach periodischen Systeme.''
\end{quote}

Nach Hamburg zur\"uckgekehrt, begann Pauli \"uber den Abschluss der
Elektronenschalen und die Beziehung zwischen diesem Problem und der
Multiplettstruktur nachzudenken. Zu diesem Zeitpunkt war man allgemein 
der Meinung, dass die Dublettstruktur der Alkalimetalle auf einem
nichtverschwindenden Drehimpuls des Atomrumpfs beruhte. Diese
Auffassung wurde von Pauli im Herbst 1924 in einer grundlegenden
Arbeit, auf die ich gleich n\"aher eingehen werde,
verworfen. Stattdessen f\"uhrte er die Annahme einer neuen
``eigent\"umlichen, klassisch nicht beschreibbaren Art von
Zweideutigkeit der quantentheoretischen Eigenschaften des
Leuchtelektrons''$^{[20]}$ ein. In einem sehr wichtigen Brief vom
24.~November 1924 an Land\'e$^{[21]}$ sagt er zu dieser
Zweideutigkeit: ``Das Leuchtelektron bringt es auf eine r\"atselhafte, 
unmechanische Weise fertig, in zwei Zust\"anden (mit gleichem $k$) mit 
verschiedenen Impulsen zu laufen.'' Pauli hat damit als erster einen
Freiheitsgrad in die Quantentheorie eingef\"uhrt, der keine
korrespondenzm\"assige Basis hat. 

\vspace{.5cm}

\noindent
{\bf 2.~Schritt: Zweiwertigkeit des Elektrons}$^{[20]}$

\medskip

\noindent
Ich m\"ochte Ihnen jetzt genauer schildern, wie er zu diesen
Schl\"ussen kam.

Im l\"angeren ersten Teil der Arbeit$^{[20]}$ (eingegangen am
2.~Dez.~1924) verwirft Pauli die \"ubliche Auffassung, nach der ``die
Elektronen der $K$-Schale an dem Zustandekommen der Komplexstruktur
und des anomalen Zeemaneffektes der optischen Spektren sehr wesentlich 
beteiligt sind''.

Dazu berechnet er zun\"achst die relativistischen Korrekturen auf das
magnetische Moment und den Bahndrehimpuls von Elektronen in der
$K$-Schale. F\"ur das Verh\"altnis der beiden findet er mit einfachen
klassischen Argumenten
\[
\frac{|\vec{M}|}{|\vec{L}|}\;=\;\frac{e}{2mc}~\big\langle \left(
  1-v^2/c^2\right)^{1/2}\big\rangle_{\rm zeitlich}~.
\]
Aufgrund des Virialsatzes ergibt sich ferner, dass der obige zeitliche 
Mittelwert gleich der Gesamtenergie des Elektrons in Einheiten $mc^2$ 
ist. F\"ur letztere verwendet er die Sommerfeldsche Formel und findet
f\"ur die $K$-Schale $(L=0, n=1)$ f\"ur den relativistischen
Korrekturfaktor den Wert $(1-\alpha^2Z^2)^{1/2}$ $[ \simeq 1
-\,\frac12~\alpha^2Z^2/n^2$~f\"ur beliebiges~$n]$.

Jetzt wird unter Annahme der `orthodoxen' Auffassung der Einfluss
der Relativit\"atskorrektur auf den anomalen Zeemaneffekt
berechnet. Dabei benutzt er seine fr\"uheren Resultate, insbesondere
die diskutierte `Summenregel'. 
Ich brauche dies nicht n\"aher auszuf\"uhren, denn es zeigt sich, dass 
die Auswirkung auf die $g$-Faktoren {\it nicht} mit den Beobachtungen
vereinbar ist. Dieses Resultat fasst Pauli so zusammen (p.~383):
\begin{quote}
``Will man an der Voraussetzung festhalten, dass auch abgeschlossene
Elektronengruppen und insbesondere die $K$-Schale der Sitz der
magneto-mechanischen Anomalie sind, so muss man nicht allein eine
Verdopplung des Quotienten aus magnetischem Moment und Impulsmoment
dieser Gruppen gegen\"uber seinem klassischen Wert, sondern ausserdem
eine Kompensation der Relativit\"atskorrektion annehmen.''
\end{quote}
Diese logische M\"oglichkeit wird von Pauli deutlich verworfen, wobei er
auch noch weitere Gr\"unde dagegen anf\"uhrt. Stattdessen schl\"agt er 
vor, dass abgeschlossene Schalen Drehimpuls Null und kein
magnetisches Moment haben. Ich zitiere (p.~385):
\begin{quote}
``Insbesondere werden bei den Alkalien die Impulswerte des Atoms und
seine Energie\"anderungen in einem \"ausseren Magnetfeld im
wesentlichen als eine alleinige Wirkung des Leuchtelektrons angesehen, 
das auch als der Sitz der magneto-mechanischen Anomalie betrachtet wird.''
\end{quote}
Und nun kommen die prophetischen Worte:
\begin{quote}
``Die Dublettstruktur der Alkalispektren, sowie die Durchbrechung des
Larmortheorems kommt gem\"ass diesem Standpunkt durch eine
eigent\"umliche, klassisch nicht beschreibbare Art von Zweideutigkeit
der quantentheoretischen Eigenschaften des Leuchtelektrons zustande.''
\end{quote}
In der gerade besprochenen Arbeit f\"uhrt er dies nicht n\"aher aus
und sagt lediglich:
\begin{quote}
``Es braucht kaum betont zu werden, dass erst die weitere Entwicklung
der Theorie zeigen muss, inwiefern eine solche Auffassung das Richtige 
trifft und ob sie weiter ausgebildet werden kann. Es stehen dieser
Auffassung grosse Schwierigkeiten entgegen, besonders im Hinblick auf
eine nat\"urliche Verbindung derselben mit dem Korrespondenzprinzip.''
\end{quote}
Ich will den zweiteiligen Schl\"usselsatz von Pauli noch etwas n\"aher
erl\"autern.

F\"ur {\it starke} Felder hatte Pauli
\[
M=M_L+M_S~,~~~ \triangle E/\mu_0B=M_L+2M_S~.
\]
Speziell f\"ur Alkaliatome tragen die abgeschlossenen Schalen nichts
zum magnetischen Moment und zu $M$ bei. ($M$ wird nach Sommerfeld als
{\it totaler} Drehimpuls in der Feldrichtung interpretiert.) Es ist
dann
\[
M=m_\ell+m_s~,~~~ \triangle E/\mu_0B = m_\ell+2m_s~,
\]
wobei $m_\ell, m_s$ die Werte von $M_L$ und $M_S$ f\"ur das
Leuchtelektron bezeichnen.

Die ganze Zahl $m_\ell$ kann klassisch als Bahndrehimpuls in der
Feldrichtung interpretiert werden. Folglich ist $m_s$ ein {\it
  intrinsischer} Beitrag des Leuchtelektrons zum totalen Drehimpuls
$M$ in Feldrichtung, der zu $m_\ell$ hinzukommt. Den beiden Termen der 
Dubletts entsprechend, hat $m_s$ die Werte $\pm \frac 12$. (Dies wurde 
bereits ausgef\"uhrt.)

Da $m_\ell$ ganz ist, muss $M$ halbganz sein, und weil $J$ als
maximaler Wert von $M$ eines Multipletts definiert ist, haben wir
f\"ur die beiden Terme eines Alkali-Dubletts 
\[
J\;=\;L\;\pm\;1/2~.
\]
Deshalb ist die Zweiwertigkeit von $J$, welche f\"ur die
Dublettaufspaltung verantwortlich ist, eine direkte Folge der
Zweiwertigkeit von $m_s$. Dies erkl\"art den ersten Teil von Paulis
Schl\"usselaussage:
\begin{quote}
``Die Dublett-Struktur der Alkalispektren \dots kommt durch eine
eigent\"umliche \dots Zwei\-deutig\-keit \dots des Elektrons zustande.''
\end{quote}

Was ist genau mit der ``Durchbrechung des Larmortheorems'' gemeint?

In der besprochenen 2.~Arbeit benutzt Pauli f\"ur die Energie\"anderung
eines Atoms in einem Magnetfeld die bekannte Formel
\[
\triangle E = - \vec{M} \cdot \vec{B}~,~~~ \vec{M}~:~{\rm magn.~Moment~.}
\]
Vergleichen wir dies mit seinem Ausdruck f\"ur starke Felder, so sehen 
wir, dass sich ein Atom in einem starken Feld wie ein Magnet mit
Moment $\mu_0 (M_L+2M_S)$ in Richtung des Feldes verh\"alt. F\"ur ein
einzelnes Valenzelektron ist dies gleich $\mu_0 (m_\ell + 2m_s)$. 

Soweit sind {\it starke} Felder vorausgesetzt. Betrachten wir aber
jetzt $S$-Zust\"ande von Alkalien, so ist $M_L=0$, $M=M_S=m_s$ und die 
Formel $\triangle E/\mu_0B = m_\ell +2m_s$ gilt aufgrund von Paulis
Summenregel auch f\"ur {\it schwache} Felder. Dies bedeutet: F\"ur
$S$-Zust\"ande ist das magnetische Moment der Alkaliatome gleich
$\underline{2m_s\mu_0}$. Dieses muss nach Pauli ausschliesslich {\it dem
  Valenzelektron zugeschrieben} werden. 

In der gleich zu besprechenden 3.~Arbeit f\"ugt er hinzu, dass er zum
formal bedingten Wert $g=2$ des Aufspaltungsfaktors beim $s$-Term der
Alkalien \dots keine n\"ahere theoretische Analyse versuchen will
(p.~767).

Pauli str\"aubte sich aus verst\"andlichen Gr\"unden zun\"achst gegen
Modellvorstellungen seines vierten Freiheitsgrades durch
Kronig$^{[22]}$, Uhlenbeck und Goudsmit$^{[23]}$. Viel sp\"ater
bemerkte er dazu$^{[17]}$: ``Obgleich ich anf\"anglich die Korrektheit
dieser Idee wegen ihres klassischen Charakters stark bezweifelte,
wurde ich schliesslich durch Thomas' Berechnung der Gr\"osse der
Dublettaufspaltung zu ihr bekehrt.''

Im erw\"ahnten Brief an Land\'e$^{[21]}$  wird auch zum ersten Mal das
Ausschliessungsprinzip formuliert. Den entscheidenden Anstoss dazu gab 
eine sehr wichtige Bemerkung in einer kurz zuvor erschienenen Arbeit
von Stoner$^{[24]}$, auf die Pauli zuf\"allig aufmerksam wurde. In
heutiger Terminologie unternahm Stoner eine Klassifikation der
Elektronen in Untergruppen zu den beiden Quantenzahlen $\ell$ und $j$
und stellte fest, dass die Anzahl der Elektronen in einer
abgeschlossenen Untergruppe zu gegebenem $\ell$ zugleich auch die Zahl 
der Zeeman-Terme in den Alkalispektren ist. In seinem
Nobelpreisvortrag$^{[17]}$  sagt Pauli dazu:
\begin{quote}
``Auf der Grundlage meiner fr\"uheren Ergebnisse \"uber die
Klassifikation der Spektralterme in einem starken magnetischen Feld
wurde mir nun die allgemeine Formulierung des Ausschliessungsprinzips
klar. Der Grundgedanke kann so ausgesprochen werden: Die komplizierten 
Anzahlen der Elektronen in abgeschlossenen Untergruppen werden auf die 
einfache Zahl Eins reduziert, wenn die Unterteilung der Gruppe durch
Angabe der Werte der 4 Quantenzahlen so weit getrieben ist, dass jede
Entartung beseitigt ist. Ein \"uberhaupt nicht entartetes
Energieniveau ist bereits `abgeschlossen', wenn es von einem
einzigen Elektron eingenommen wird; Zust\"ande, die diesem Postulat
widersprechen, sind auszuschliessen.''
\end{quote}

Ich m\"ochte auch hier etwas in die Einzelheiten gehen.

In seiner entscheidenden dritten Arbeit$^{[14]}$  fasst Pauli zun\"achst
seine bisherigen Ergebnisse f\"ur die Alkalien zusammen
(p.~766). F\"ur diese fallen die Quantenzahlen $L,J,M$ des Atoms mit
denjenigen des Leuchtelektrons zusammen, f\"ur die wir die modernen
Bezeichnungen $\ell, j, m_j$ verwenden. [Paulis Notation ist:
~$k_1=\ell+1,~ k_2=j+\,\frac 12 ~, m_1=m_j$.] Daneben gibt es noch die 
Hauptquantenzahl $n$. Wie bereits ausgef\"uhrt, ist bei den Alkalien
$j=\ell \,\pm\,\frac 1 2~.$ ~Pauli betont:
\begin{quote}
``Die Anzahl der Zust\"ande im Magnetfeld bei gegebenen $\ell$ und $j$ 
ist $2j+1$, die Anzahl dieser Zust\"ande f\"ur beide Dubletts mit
gegebenem $\ell$ zusammengenommen ist ~$\underline{2(2\ell+1)}$.''
\end{quote}
Nun geht Pauli daran, die ``formale Klassifikation des Leuchtelektrons
durch die vier Quantenzahlen $n, \ell, j, m_j$ 
auch auf {\it komplizierte Atome} zu \"ubertragen.'' Dabei beruft er
sich auf {\it Bohrs   Permanenzprinzip (Aufbauprinzip)}, welches
folgendes besagt: Wird an ein teilweise ionisiertes Atom ein weiteres
Elektron angelagert, dann behalten die Quantenzahlen der schon
gebundenen Elektronen Werte, die ihnen im ionisierten Zustand
zukommen. Pauli zeigt, dass damit in einfachen, aber auch in
komplizierten F\"allen die richtige Termmannigfaltigkeit
herauskommt. (Dabei beruft er sich auf eine Verzweigungsregel von
Heisenberg und Land\'e.)

F\"ur den weiteren Gedankengang sind nun wieder die Formeln f\"ur den
Zeemaneffekt in {\it starken} Feldern wesentlich. Zun\"achst
impliziert das Permanenzprinzip f\"ur ein Atom in einem starken Feld
die Existenz von Quantenzahlen $m_j$ f\"ur die einzelnen Elektronen,
deren Summe gleich dem totalen Drehimpuls $M$ des Atoms in
Feldrichtung ist:
\[
M\;=\;\sum m_j~.
\]
Ferner setzt sich nach der Permanenzregel auch das magnetische Moment
$(M+M_S)\mu_0$ ebenfalls additiv aus Momenten $m_2\mu_0$ der einzelnen 
Elektronen zusammen:
\[
M_2 := M_L + 2M_S = M+M_S = \sum m_2,~ m_2=m_j\;\pm\;1/2~.
\]
In den Summen sollten unabh\"angig voneinander alle Werte durchlaufen
werden, welche zu den Quantenzahlen $j,\ell$ geh\"oren.

Nach diesen Vorbereitungen sieht sich Pauli zur folgenden grundlegenden
Annahme aufgefordert:

\begin{quote}
``Jedes Elektron im Atom kann durch eine Hauptquantenzahl $n$ und
  drei zus\"atzliche Quantenzahlen $\ell, j, m_j$ charakterisiert werden.''
\end{quote}
Es ist, wie bei den Alkalispektren, $j=\ell\pm 1/2$. An Stelle von $j$ 
k\"onnen wir auch $m_2=m_j\pm 1/2$ verwenden. 

Nun betrachtet Pauli den Fall {\it \"aquivalenter Elektronen}. Zun\"achst 
bemerkt er, dass in diesem Fall gewisse Kombinationen der
Quantenzahlen in der Natur {\it nicht} vorkommen. Falls beispielsweise 
zwei Valenzelektronen in $s$-Zust\"anden zu verschiedenen
Hauptquantenzahlen geh\"oren, gibt es sowohl einen Singlett-$S$-Term
als auch einen Triplett-$S$-Term; wenn aber beide Elektronen dasselbe
$n$ haben, so kommt {\it nur der Singlett-Term} vor. F\"ur Pauli stellt
sich ``die Frage, durch welche quantentheoretischen Regeln dieses
Verhalten der Terme beherrscht wird'' (p.~772).

Dazu sagt er gleich, dass diese Frage aufs engste mit dem Problem der
abgeschlossenen Schalen verkn\"upft ist. Dazu hatte k\"urzlich
E.C. Stoner einen neuen, von Bohr abweichenden  Vorschlag
gemacht. Bohr hatte in seiner Theorie des periodischen Systems
beispielsweise die 8 Elektronen der $L$-Schale in zwei
4er-Untergruppen unterteilt. Stoner schlug hingegen vor, diese in eine 
Untergruppe von 2 Elektronen mit $\ell=0$ und eine Untergruppe von 6
Elektronen mit $\ell =1$ aufzuteilen. Allgemein ordnete er jedem Wert
von $\ell < n$ eine Untergruppe von $2(2\ell+1)$ Elektronen zu.

Wichtiger noch war Stoners Bemerkung, dass dieselbe Zahl $2(2\ell+1)$
auch mit der Anzahl der Zust\"ande eines {\it Alkaliatoms in einem
  Magnetfeld} mit demselben Wert von $\ell$ und einem gegebenen Wert
der Hauptquantenzahl des Valenzelektrons \"ubereinstimmt. 

Diese Bemerkung war f\"ur Pauli der Schl\"ussel zu seinem
Ausschlussprinzip: Er erkl\"art die Tatsache, dass es f\"ur jede
Untergruppe zu gegebenem $(n, \ell)$ einer abgeschlossenen Schale
genau $2(2\ell+1)$ Elektronen gibt, durch die Annahme, dass jeder
Zustand zu den charakteristischen Quantenzahlen $(n, \ell, j, m_j)$
{\it genau durch ein Elektron besetzt} ist. Wir haben dann n\"amlich,
wie bereits bemerkt, die beiden M\"oglichkeiten $j=\ell \pm \frac 1 2$ 
mit $2j+1$ Werten f\"ur $m_j$, was insgesamt $2(2\ell+1)$
M\"oglichkeiten ergibt.

Es muss hier nochmals betont werden, dass Pauli die Quantenzahlen $j$
und $m_j$ f\"ur einzelne Elektronen in starken Feldern definiert
hatte. Dazu bemerkt er (p.~776):
\begin{quote}
``Aus thermodynamischen Gr\"unden [Invarianz der statistischen
Gewichte bei adiabatischen Transformationen des Systems] muss jedoch
die Anzahl der station\"aren Zust\"ande des Atoms bei gegebenen Werten 
der Zahlen $k_1$ und $k_2$ f\"ur die einzelnen Elektronen und des
Wertes von $\overline{m}_1 = \sum m_1$ f\"ur das ganze Atom in
starken und in schwachen Feldern \"ubereinstimmen.''
\end{quote} 

Anf\"anglich war Pauli offenbar nicht sicher, wie weit sich sein
Prinzip bew\"ahren w\"urde, bezeichnete er es doch in seinem Brief an
Bohr$^{[25]}$  als ``Unsinn \dots, der zum bisher \"ublichen Unsinn
konjugiert'' ist. Durch das Ausschliessungsprinzip wurden aber
sogleich viele Tatsachen der Atomphysik verst\"andlich, so nat\"urlich 
die Periodenl\"angen 2, 8, 18, 32, \dots der Elemente. Pauli
beschliesst seine ber\"uhmte Arbeit$^{[14]}$ vom 16.~Januar 1925 mit
den Worten: 
\begin{quote}
``Das Problem der n\"aheren Begr\"undung der hier zugrunde
gelegten allgemeinen Regel \"uber das Vorkommen von \"aquivalenten
Elektronen im Atom d\"urfte wohl erst nach einer weiteren Vertiefung
der Grundprinzipien der Quantentheorie erfolgreich angreifbar sein.'' 
\end{quote}

In Anbetracht der anf\"anglich negativen Einstellung zur Idee eines
Elektronenspins ist es eigentlich erstaunlich, dass Pauli noch {\it
  vor} dem Ausschliessungsprinzip die Vorstellung eines
Kerndrehimpulses einf\"uhrte, um gewisse `Satelliten einiger
Spektrallinien' zu deuten$^{[26]}$. Dieser Mechanismus wurde sp\"ater 
als Hyperfein-Wechselwirkung bezeichnet. 

Klassisch, und immer noch sehr sch\"on zu lesen, ist Paulis Arbeit
`Zur Quantenmechanik des magnetischen Elektrons' aus dem Jahre 1927, 
in welcher er mit Hilfe der nach ihm benannten Spinmatrizen die
nichtrelativistische Wellenmechanik von Spin 1/2-Teilchen endg\"ultig
formulierte$^{[27]}$. Die Beschreibung eines Zustandes von endlich
vielen Elektronen durch eine mehrkomponentige Wellenfunktion, welche
sich nach einer zweideutigen Darstellung der Rotationsgruppe
transformiert, war auch im Hinblick auf die relativistische
Dirac-Gleichung grundlegend $^{[22]}$.

Die grosse Bedeutung des Pauliprinzips --- weit \"uber die Atomphysik
hinaus --- hat Ehrenfest in seiner bereits erw\"ahnten Ansprache zur
Verleihung der Lorentzmedaille an Pauli in sehr lebendiger und
eing\"angiger Weise erl\"autert$^{[2]}$. Treffend sagt er z.B.: 
``Darum sind also die Atome so unn\"otig dick; darum der Stein, das
Metallst\"uck etc. so volumin\"os! Sie m\"ussen zugeben, Herr Pauli: 
Durch eine {\it partielle} Aufhebung Ihres Verbotes k\"onnten Sie uns
von gar vielen Sorgen des Alltags befreien: z.~B. vom Verkehrsproblem
unserer Strassen.'' ~Tats\"achlich ist das Pauli-Prinzip entscheidend
f\"ur die Stabilit\"at der makroskopischen Materie. Wir m\"ussen
deshalb --- mit den Worten von Ehrenfest$^{[2]}$ --- ``jedesmal, wenn
ein Atomkern bei einem Beta-Zerfall ein neues Elektron in die Welt
setzt, mit Herzklopfen abwarten; wird sich nun das neue Elektron
gehorsam dem Pauli-Verbot f\"ugen oder wird es in boshaftem \"Ubermut
den Anti-Symmetrie-Tanz seiner \"alteren Geschwister verwirren?''

Auch auf die Evolution und die Struktur der Sterne --- ganz besonders
auf deren Endzust\"ande --- hat das Ausschliessungsprinzip
fundamentale Auswirkungen. Es ist dabei sehr bemerkenswert, dass die
Quantenstatistik identischer Teilchen ihre erste Anwendung in der 
Astrophysik fand. Bereits Ende 1926 entwickelte Fowler auf ihrer Basis 
die (nichtrelativistische) Theorie der Weissen Zwerge. 

\vspace{.5cm}

\section*{Quantenmechanik}

Nach der mathematischen Ausgestaltung der Matrizenmechanik im Jahre
1925 stellte sich das dringende Problem, das Wasserstoffspektrum aus
der neuen Theorie herzuleiten. In diesem Stadium, als die
Wellenmechanik noch nicht vorlag, war dies eine schwierige Aufgabe,
um deren L\"osung sich z.~B. Heisenberg vergeblich bem\"uhte. Schon im
Oktober 1925 hatte Pauli aber die vollst\"andige Quantenmechanik des
Wasserstoffatoms. Auf das brieflich mitgeteilte Resultat schrieb
Heisenberg$^{[28]}$: ``Ich brauche Ihnen wohl nicht zu schreiben, wie
sehr ich mich \"uber die neue Theorie des Wasserstoffs freue, und wie
sehr ich es bewundere, wie Sie diese Theorie so schnell herausgebracht 
haben.'' Pauli deckte bei seiner raffinierten L\"osung eine verborgene 
vierdimensionale Drehsymmetrie auf, welche es ihm gestattete, das
Wasserstoffspektrum auf rein algebraischem Weg zu berechnen. Eine
wichtige Rolle spielte dabei der von seinem damaligen Chef in Hamburg
in die Atomphysik eingef\"uhrte Lenzsche Vektor (dessen Urspr\"unge
freilich weit zur\"uckreichen).

Nachdem im Fr\"uhjahr des n\"achsten Jahres durch Schr\"odinger die
vollst\"andige mathematische Formulierung der Wellenmechanik
entstanden war, wurden die Interpretationsfragen der Quantenmechanik
immer dr\"angender. Die wichtige Rolle, die Pauli in den kommenden
Auseinandersetzungen gespielt hat, wurde von Heisenberg im
Erinnerungsband f\"ur Wolfgang Pauli lebendig
beschrieben$^{[29]}$. Leider sind aus dieser Zeit alle Briefe von
Pauli an Heisenberg verschollen --- bis auf ein sehr aufschlussreiches 
Schreiben vom 19.~Oktober 1926, welches das gefl\"ugelte Wort
enth\"alt$^{[30]}$: ``Man kann die Welt mit dem $p$-Auge und man kann
sie mit dem $q$-Auge ansehen, aber wenn man beide Augen zugleich
aufmachen will, dann wird man irre.'' Pauli setzt im gleichen Brief
--- neben dieser Andeutung der Unsch\"arferelation --- auch
auseinander, dass die Bornsche Interpretation nur ein Spezialfall einer 
viel allgemeineren Interpretationsvorschrift sei. Beispielsweise sei
$|\tilde{\Psi}(p)|^2$ die Wahrscheinlichkeitsdichte 
im Impulsraum. Der Brief enth\"alt ausserdem eine Erkl\"arung des
Paramagnetismus, bei der Pauli erstmals die Quantenstatistik in der
Gibbs'schen grosskanonischen Gesamtheit entwickelte. 

In einem sehr bemerkenswerten Brief an Jordan$^{[31]}$ erbrachte Pauli 
--- unabh\"angig von Schr\"odinger --- den Nachweis der \"Aquivalenz
von Matrizenmechanik und Wellenmechanik. 

Pauli hatte grossen Anteil an der `Kopenhagener Interpretation' der
Quantenmechanik und war vielleicht deren konsequentester Verfechter. Er 
ist auch sp\"ater wiederholt auf dieses Thema zur\"uckgekommen.

W\"ahrend der ganzen Entstehungszeit der Quantenmechanik hatte Pauli
die Pr\"uf- und Kontrollfunktion. ``Bei ihm lief die Information ein,
hier wurde sie entwickelt, von hier strahlte sie gereinigt wieder
aus. Daher sind am Briefwechsel alle beteiligt, die mitgestaltet haben 
--- bis auf den grossen Einsamen, der sich selbst gen\"ugte: Paul
Adrian Maurice Dirac.''$^{[32]}$

\vspace{.5cm}

\section*{Quantenelektrodynamik, Quantenfeldtheorie}

Nach Vollendung der nicht-relativistischen Quantenmechanik galt Paulis 
Hauptinteresse der Quantenelektrodynamik und Quantenfeldtheorie, welche 
er in entscheidender Weise gef\"ordert hat.

Bereits in der grossen `Dreim\"annerarbeit' erkannten Born,
Heisenberg und Jordan, dass die Neuinterpretation der physikalischen
Observablen auch auf das Maxwell-Feld ausgedehnt werden musste. Da das 
freie elektromagnetische Feld als ein System von unendlich vielen
ungekoppelten harmonischen Oszillatoren aufgefasst werden kann, war
klar, wie die Quantisierung zu geschehen hatte. Dabei tauchten im
Lichtwellenfeld ganz von selber die korpuskularen Teilchen auf, die
Einstein in raffinierten statistischen \"Uberlegungen aus dem
Planckschen Gesetz herausgelesen hatte. Die Feldquanten gehorchten
dabei der Bose-Einstein Statistik.

Im Jahre 1927 behandelte Dirac als erster das quantisierte
elektromagnetische Feld in Wechselwirkung mit einem materiellen
System, welches durch die nichtrelativistische Quantenmechanik
beschrieben wurde. Erfolg und Misserfolg der Theorie waren sofort
klar. Dirac leitete aus seiner Theorie konsequent die
Einstein-Koeffizienten f\"ur Emission und Absorption von Licht
her. Ehrenfest wies aber sogleich darauf hin, dass die Theorie in
h\"oheren Ordnungen der St\"orungstheorie zu Divergenzen f\"uhren
m\"usse, da in ihr das Vektorpotential am Ort des Elektrons eingeht.

In einer sehr wichtigen Arbeit von Jordan und Pauli$^{[33]}$, `Zur
Quantenelektrodynamik ladungsfreier Felder' wird erstmals an der
manifesten Lorentzinvarianz bei der Quantisierung festgehalten und es
werden die lorentzinvarianten Vertauschungsrelationen f\"ur die
elektromagnetischen Feldst\"arken abgeleitet. Merkw\"urdigerweise wird 
in dieser Arbeit noch nicht betont, dass die Kommutatoren der
Feldst\"arken f\"ur raumartige Separationen verschwinden, 
was nat\"urlich im Hinblick auf die Beziehung von Spin und Statistik
sehr wesentlich ist.

Da die Anwendung der beim Strahlungsfeld benutzten
Quantisierungsvorschriften auf das Schr\"odingersche Wellenfeld
ebenfalls zur Bose-Einstein Statistik f\"uhrt, ersetzten Jordan und
Wigner die Kommutatoren f\"ur die Fourierkoeffizienten des Feldes
durch Antikommutatoren. 

Im Anschluss an diese fr\"uhen Entwicklungen entstehen nun die zwei
grossen Arbeiten von Heisenberg und Pauli$^{[34]}$, in denen zum
ersten Mal eine systematische Theorie der Feldquantisierung entwickelt 
wird. Da die kanonische Quantisierung die Zeit auszeichnet und daher
von Natur aus nicht manifest lorentzinvariant ist, ist der Nachweis
der Lorentzinvarianz des Quantisierungsverfahrens ein
Hauptanliegen. Dieser erweist sich als ziemlich umst\"andlich;
trotzdem halten aber die Autoren aus verschiedenen Gr\"unden am
dreidimensionalen Standpunkt fest. Das wichtigste Modell einer
quantisierten Feldtheorie --- die Quantenelektrodynamik --- wird von
ihnen in den allgemeinen Rahmen eingebaut.

Damit waren die Grundlagen gelegt, auf denen alle weiteren
Entwicklungen der Quantenfeldtheorie und insbesondere der
Quantenelektrodynamik aufbauten. Ich beschr\"anke mich hier auf eine
Skizze der wichtigsten Etappen, die Pauli schliesslich 1939 zu seinem
eleganten Beweis f\"ur den {\it Zusammenhang zwischen Spin und
  Statistik} f\"uhrten, womit er f\"ur sein Ausschliessungsprinzip
eine tiefere Begr\"undung fand. 

Paulis anf\"angliche Abneigung gegen Diracs k\"uhne L\"ochertheorie,
in welcher das Ausschliessungsprinzip zur Stabilisierung des Vakuums
benutzt wurde, sind bekannt und finden z.~B. im Handbuchartikel von
1933 --- noch kurz vor der Entdeckung des Positrons durch
C.D. Anderson --- ihren beredten Niederschlag$^{[35]}$. Deshalb nannte 
er auch eine gemeinsame Arbeit mit Weisskopf, auf die ich nun zu
sprechen komme, f\"ur l\"angere Zeit die `Anti-Dirac-Theorie'.

Pauli und Weisskopf studieren in ihrer Arbeit$^{[36]}$ die zum
Schulbeispiel gewordene Quantisierung eines komplexen Skalarfeldes
$\psi (x)$, das wegen des Bestehens einer Eichgruppe nicht beobachtbar 
ist. Deshalb darf das Lokalisierungspostulat nicht direkt auf das
$\psi$-Feld angewandt werden, wohl aber auf den erhaltenen Strom, der
bilinear aus $\psi$ und $\psi^*$ aufgebaut ist. Dieses ist erf\"ullt,
falls $\psi(x)$ mit $\psi(x')$ kommutiert und $\psi(x)$ und
$\psi^*(y)$ f\"ur raumartige Separationen der Argumente
kommutieren. Nun w\"urde man auch einen lokalen Stromoperator
erhalten, falls $\psi(x)$ und $\psi(x')$ antikommutiert und der
Antikommutator von $\psi(x)$ und $\psi^*(x)$ f\"ur raumartige
Abst\"ande verschwindet. Diese letzte Forderung impliziert aber ---
zumindest im kr\"aftefreien Fall --- dass $\psi$ identisch
verschwindet. Deshalb ist nur eine Quantisierung m\"oglich, bei der
die Teilchen der Bose-Einstein Statistik gen\"ugen.

Pauli teilte dieses Ergebnis in einem Brief an Heisenberg$^{[37]}$ mit 
folgenden Worten mit:
\begin{quote}
``Die Frage, ob sich  bei dieser Theorie formal auch eine Quantelung
der Materiewellen mit Ausschliessungsprinzip durchf\"uhren l\"asst,
bedurfte einer genaueren Untersuchung. Das Ergebnis ist
interessanterweise negativ.''
\end{quote}

Um im Falle des Diracfeldes ein stabiles Vakuum zu erhalten, muss mit
dem Formalismus von Jordan und Wigner nach dem Ausschlussprinzip
quantisiert werden. Dann erh\"alt man in symmetrischer Weise Teilchen
beider Ladungen.

Pauli war nun bald \"uberzeugt, dass ``hier ein zwangsl\"aufiger
Zusammenhang von Spin und Statistik aufzud\"ammern beginnt''$^{[38]}$. 
Bevor ein allgemeiner Beweis m\"oglich wurde, musste aber zuerst die
Theorie kr\"aftefreier Felder zu beliebigem Spin entwickelt
werden. Dies leistete --- nach Vorarbeiten von Dirac$^{[39]}$ ---
Markus Fierz$^{[40]}$, der zu jener Zeit Assistent bei Pauli war. Die
bedeutende Abhandlung von Fierz ist bei der j\"ungeren Generation
leider etwas in Vergessenheit geraten, wohl haupts\"achlich deshalb,
weil Pauli in seiner ber\"uhmten Arbeit mit dem Titel `The Connection
Between Spin and Statistics'$^{[41]}$ das Fierzsche Hauptresultat
f\"ur nicht notwendigerweise irreduzible Spinorfelder ohne viel
Rechnung auf sehr elegante Weise beweisen konnte. Fierz und Pauli
zeigen beide, dass in einer bez\"uglich der eigentlichen Lorentzgruppe 
invarianten ($c$-Zahl-)Theorie freier Felder {\it die Ladungsdichte
  f\"ur eindeutige und die Energiedichte f\"ur zweideutige
  Darstellungen indefinit sind.} Um im Falle zweideutiger Felder
(halbzahliger Spin) ein stabiles Vakuum zu erhalten, muss man deshalb
nach dem Ausschlussprinzip quantisieren. Dann erh\"alt man, wie bei
der Quantisierung des Diracfeldes, Teilchen beider Ladungen in
symmetrischer Weise. F\"ur den Fall eindeutiger Felder (ganzzahliger
Spin) ist die Quantisierung nach Bose-Einstein m\"oglich. Hingegen ist 
die Quantisierung nach dem Ausschlussprinzip nicht mit einer lokalen
Ladungsdichte vertr\"aglich. Mit Recht beschliesst Pauli seine Arbeit
mit folgende Worten:
\begin{quote}
``In conclusion we wish to state that according to our opinion the
connection between spin and statistics is one of the most important
applications of the special relativity theory.''
\end{quote}

\vspace{.5cm}

\section*{Neutrinohypothese}

Es ist heute nur noch schwer vorstellbar, welcher Mut im Jahre 1930
n\"otig war, um die Existenz eines neuen leichten elektrisch neutralen 
Spin 1/2-Teilchens zu postulieren, welches sich bis anhin allen
Beobachtungen entzogen hatte. Die damalige Physik kannte schon seit
geraumer Zeit nur drei Elementarteilchen: Das unabh\"angig von
Wiechert und Thompson entdeckte Elektron, das Proton als leichtester
Atomkern und das Photon, welches Einstein mit Hilfe des Boltzmannschen 
Prinzips aus dem asymptotisch g\"ultigen Wienschen Gesetz f\"ur die
Hohlraumstrahlung herauskristallisiert hatte. 

Paulis Neutrinohypothese war anf\"anglich mit gewissen Schwierigkeiten
verbunden, weil er zwei g\"anzlich verschiedene Probleme der damaligen 
Kernphysik auf einen Schlag l\"osen wollte. Das eine Problem betraf
das 1914 von Chadwick entdeckte kontinuierliche Spektrum beim
Beta-Zerfall und das andere die `verkehrte' Statistik der Atomkerne. 
Da letzteres sp\"ater --- nach der Entdeckung des Neutrons durch
Chadwick im Jahre 1932 -- wegfiel, muss kurz auf das damalige Problem
von Spin und Statistik der Atomkerne eingegangen werden. 

Nach Rutherfords Versuchen \"uber k\"unstliche Kernumwandlungen hatte
man angenommen, dass die Kerne aus Protonen und Elektronen
best\"unden. Dann musste aber z.~B. der Stickstoffkern mit der
Massenzahl 14 und der Ladung 7 aus 21 Teilchen bestehen, n\"amlich 14
Protonen und 7 Elektronen. Deshalb sollte der Stickstoffkern
halbzahligen Spin haben und somit nach dem `Wechselsatz' dem
Auschliessungsprinzip (Fermi-Statistik) gen\"ugen. 
Durch Analyse der Bandenspektren von N$_2$-Molek\"ulen zeigte
Kronig$^{[42]}$, dass der Stickstoff-Kern Spin 1 hat und Heitler und
Herzberg$^{[43]}$ kamen zum Schluss, dass er die Bose-Statistik
befolgt.

Aus den diskreten Alpha- und Gamma-Spektren musste man schon
fr\"uhzeitig schliessen, dass auch die Kernzust\"ande diskret sind und 
deshalb war die Deutung des kontinuierlichen Beta-Spektrums sehr
r\"atselhaft. Lise Meitner u.a. stellten sich zun\"achst vor, dass die
diskreten Prim\"arenergien durch Sekund\"arprozesse in ein Kontinuum
verbreitert werden. Man konnte sich beispielsweise denken, dass ein
Prim\"arelektron an den Elektronen der \"ausseren Schalen gestreut
wird, oder Elektronen geringerer Energie emittiert. Durch kalorische
Pr\"azisionsmessungen von L. Meitner$^{[44]}$ wurde aber im Jahre 1930 
schliesslich sicher gestellt, dass das kontinuierliche Spektrum den
Prim\"arelektronen zuzuschreiben ist.

Pauli hat seinen `verzweifelten Ausweg' zuerst in einem ber\"uhmten
`offenen Brief' vom 4.~Dezember 1930$^{[45]}$, der von Lise Meitner
aufbewahrt wurde, zur Diskussion gestellt. Weniger bekannt ist ein
Brief an Oskar Klein$^{[46]}$ vom 12.~Dezember 1930, in welchem er
n\"aher auf die Eigenschaften des hypothetischen `Neutrons' eingeht,
das sp\"ater von Fermi in Neutrino umgetauft wurde. Es wird darin auch 
auseinandergesetzt, warum Pauli --- in scharfem Gegensatz zu Bohr ---
an der uneingeschr\"ankten G\"ultigkeit des Energiesatzes festhalten
wollte. Dieser Brief an O. Klein ist derart aufschlussreich, dass ich
hier einen l\"angeren Teil zitiere:

\begin{quote}
``Anl\"asslich der L\"osung einer Schulaufgabe \"uber
Hyperfeinstruktur von Li --- welche L\"osung Du in Form einer
gemeinsamen Arbeit von Herrn G\"uttinger und mir demn\"achst in der
Zeitschrift f\"ur Physik finden wirst --- habe ich mir \"uber die
`verkehrte' Statistik der Kerne sowie \"uber das kontinuierliche
$\beta$-Spektrum noch einmal gr\"undlich den Kopf zerbrochen. Dann
fiel mir folgender m\"oglicher Ausweg ein (ein Ausweg der Verzweiflung 
allerdings): Es k\"onnten die Kerne ausser Elektronen und Protonen
noch andere Elementarteilchen enthalten und zwar m\"ussten diese
elektrisch neutral sein, der Fermi-Statistik gehorchen und den Spin
1/2 haben. Nennen wir diese Teilchen {\it Neutronen}. Es ist klar,
dass man dann die `verkehrte' Statistik verstehen kann. Auch die
$\beta$-Spektren k\"onnte man dann mit Beibehaltung des Energiesatzes
verstehen, wenn man annimmt, dass beim $\beta$-Zerfall ausser dem
Elektron immer noch ein Neutron ausgeschleudert wird. Aus den
Atomgewichtsbestimmungen von Anfangs- und Endprodukten radioaktiver
Zerfallsreihen m\"usste man dann schliessen, dass die Neutronenmasse
nicht gr\"osser sein kann als 0,01 Protonmasse; sie {\it k\"onnte}
aber demnach gr\"osser sein als die Elektronenmasse. --- Nun kommt es
aber wesentlich darauf an, welche Kr\"afte auf diese Neutronen
wirken. W\"urden \"uberhaupt keine oder zu schwache Kr\"afte auf sie
wirken, so k\"onnten sie ja gar nicht im Kern bleiben. Das vom
Standpunkt der Diracschen Theorie fast einzig m\"ogliche Modell des
Neutrons w\"are dieses, dass es in einem \"ausseren Feld $F_{\mu\nu}$
der Wellengleichung
\[
\gamma^\mu~\frac{h}{2\pi i}~\frac{\partial\psi}{\partial x_\mu}\;+\;
\mu (i \gamma^\mu \gamma^\nu)\;F_{\mu\nu}\psi\,-\,imc\psi\;=\;0
\]
$\phantom{mmmmmmmmmmmm}$ (wie \"ublich~ $\gamma^\mu
\gamma^\nu\,+\,\gamma^\nu \gamma^\mu\;=\;2\delta_{\mu\nu}$)
\medskip

\noindent
gen\"ugt. Wegen $e=0$ sind hierin die Terme mit dem Potential
gestrichen; ferner {\it folgt} f\"ur solche Teilchen, dass sie sich
f\"ur langsame Geschwindigkeiten wie ein magnetischer Dipol vom Moment 
$\mu$ verhalten (die Konstante $\mu$ hat die Dimen\-sion Ladung mal
L\"ange). --- Aber das hat alles einen grossen Haken. Nimmt man
n\"amlich f\"ur $\mu$ ein gew\"ohnliches Magneton an, so w\"are, wie
eine einfache Absch\"atzung zeigt, die ionisierende Wirkung der
Neutronen nicht wesentlich kleiner als die der $\beta$-Teilchen und
alle Wilsonaufnahmen m\"ussten geradezu wimmeln von Neutronen. Selbst
wenn man $\mu$ von der Gr\"ossenordnung des Protonenmagnetons und $m$
so gross wie m\"oglich (0,01 Protonmasse) ansetzt, wird die
ionisierende Wirkung voraussichtlich nicht gr\"ossenordnungsm\"assig
kleiner als die der $\gamma$-Strahlen resultieren. Wenn die Neutronen
also wirklich existieren w\"urden, w\"are es wohl kaum verst\"andlich, 
dass man sie noch nicht beobachtet hat. Deshalb glaube ich auch selber 
nicht so ganz an die Neutronen, habe nichts dar\"uber publiziert und
habe nur einige Experimentalphysiker veranlasst, nach durchdringlichen 
Teilchen dieser Art besonders zu suchen. Ich w\"urde sehr gerne
wissen, was Bohr dazu meint.

\centerline{---------------------}

\"Uberschrift f\"ur das folgende: `Nicht um zu kritisieren, nur um zu 
verstehen!' ~In dieser Verbindung habe ich ferner f\"ur den Fall, dass 
die Neutronenidee sich als falsch herausstellen sollte, \"uber die
M\"oglichkeit eines Versagens des Energiesatzes f\"ur die Elektronen
im Kern nachgedacht und m\"ochte gerne auf dem Wege \"uber Dich, Bohr
um einige Ausk\"unfte dar\"uber bitten. Ich kann mich vorl\"aufig
nicht entschliessen, an ein Versagen des Energiesatzes ernstlich zu
glauben und zwar aus folgenden Gr\"unden (von denen ich nat\"urlich
zugebe, dass sie nicht {\it absolut} zwingend sind). 

Erstens scheint es mir, dass der Erhaltungssatz f\"ur Energie-Impuls
dem f\"ur die Ladung doch sehr weitgehend analog ist und ich kann
keinen theoretischen Grund daf\"ur sehen, warum letzterer noch gelten
sollte (wie wir es ja empirisch f\"ur den $\beta$-Zerfall wissen),
wenn ersterer versagt. Zweitens m\"usste bei einer Verletzung des
Energiesatzes auch mit dem {\it Gewicht} etwas sehr Merkw\"urdiges
passieren. Denke Dir einen geschlossenen Kasten, in welchem
$\beta$-Strahler radioaktiv zerfallen; die $\beta$-Teilchen m\"ogen
dann irgendwie an der Wand absorbiert werden und den Kasten nicht
verlassen k\"onnen. Einzelbeobachtungen dar\"uber, was im Kasten vor
sich geht, m\"ogen nicht gemacht werden, es m\"oge nur das
Gesamtgewicht des Kastens (beliebig genau) festgestellt werden. Wenn
dann der Energiesatz beim $\beta$-Zerfall nicht gelten w\"urde,
m\"usste das Gesamtgewicht des geschlossenen Kastens sich dabei
\"andern (dieser Schluss scheint mir ganz zwingend). Dies widerstrebt
meinem physikalischen Gef\"uhl auf das \"ausserste! Denn es muss dann
sogar auch f\"ur das Gravitationsfeld, das von dem ganzen Kasten (samt 
seinem radioaktiven Inhalt) selber erzeugt wird (dass man dieses wegen 
seiner Kleinheit {\it praktisch} nicht messen kann, tut nichts zur
Sache), angenommen werden, dass es sich \"andern kann, w\"ahrend wegen 
der Erhaltung der Ladung das nach aussen erzeugte elektrostatische
Feld (beide Felder scheinen mir doch analog zu sein; das wirst Du ja
\"ubrigens auch aus Deiner f\"unfdimensionalen Vergangenheit noch
wissen) unver\"andert bleiben soll. Ich m\"ochte Bohr doch ernstlich
fragen, ob er das glaubt, bzw. wie er das plausibel machen kann! Ich
w\"are ihm also {\it sehr} dankbar, wenn ich bald von ihm einen Brief
dar\"uber bekommen w\"urde, der mit dem Satz beginnt: ``Wir sind ja
ganz einig, aber \dots''~.''
\end{quote}

Pauli hat wiederholt \"offentlich \"uber seine Idee neuer sehr
durchdringender Teilchen berichtet, aber l\"angere Zeit nichts
drucken lassen. Erst am Solvay-Kongress \"uber Atomkerne im Oktober
1933 legte er seine Zur\"uckhaltung ab, da nun nach der Entdeckung des 
Neutrons eine weitgehende Kl\"arung erfolgt war. Den kurzen
Diskussionsbeitrag seiner Ideen \"uber das Neutrino liess er im
Kongressbericht drucken. 

\medskip

In dieser Skizze musste ich mich auf Paulis wichtigste Beitr\"age 
zur Physik beschr\"anken.
In sp\"ateren Jahren verfolgte er aktiv die interessanten
Fortschritte in der Quantenelektrodynamik und in der
Elementarteilchenphysik. Durch eigene Beitr\"age und die Kritik der
Arbeit anderer Forscher hat er die weitere Entwicklung entscheidend
gef\"ordert. Besonders wichtig war sein Beitrag zur CPT-Symmetrie, die 
im wesentlichen von J. Schwinger und G. L\"uders entdeckt wurde, aber
erst durch Pauli die endg\"ultige und allgemeine Formulierung
erhielt$^{[47]}$. Res Jost zeigte daraufhin, dass diese wichtige
verborgene Symmetrie ganz allgemein im Kausalit\"atspostulat der
Relativit\"atstheorie verankert werden kann$^{[48]}$. Bekannt
geblieben aus dieser Zeit sind das nach Pauli und Villars benannte
Regularisierungsverfahren$^{[49]}$ in der Feldtheorie 
und die Pauli-Gruppe, die Pauli bei seiner Diskussion der
Leptonzahlerhaltung im Jahre 1957 einf\"uhrte. 

Noch wenig bekannt ist, dass Pauli bereits 1953 wesentliche Teile der
nicht-Abelschen Eichtheorien entwickelt hat (die ber\"uhmte Arbeit von 
Yang und Mills stammt aus dem Jahre 1954). In detaillierten Briefen an 
A. Pais$^{[50]}$ und in zwei Seminarvortr\"agen in Z\"urich
entwickelte er eine nicht-Abelsche Kaluza-Klein-Theorie und die
zugeh\"orige Differentialgeometrie. Durch Spezialisierung
(Unabh\"angigkeit der Spinorfelder vom internen Raum) erhielt er alle
wichtigen Formeln von Yang und Mills, wie er sp\"ater in einem Brief
an Yang$^{[51]}$ n\"aher ausf\"uhrte. Paulis Sch\"uler P. Gulmanelli
hat die erw\"ahnten Seminarvortr\"age vom 16. und 23.~November 1953
ausgearbeitet$^{[52]}$. Daraus geht u.a. hervor, dass auch bereits die 
Yang-Mills-Wirkung n\"aher diskutiert wurde.

Leider hat Pauli \"uber seine Untersuchungen nichts publiziert, da er
der Meinung war, dass die Eichbosonen unausweichlich masselos sein
m\"ussten. (Siehe dazu auch Ref.$^{[53]}$.)

\medskip

\centerline{\Large *~~~*~~~*~~~*~~~*}

\vspace{.5cm}

Lassen Sie mich diesen Vortrag mit zwei Zitaten beschliessen. Kurz
nachdem Pauli den Nobelpreis erhalten hatte, schrieb Hermann
Weyl$^{[54]}$:
\begin{quote}
``The mathematicians feel near to Pauli since he is distinguished
among physicists by his highly developed organ for mathematics. Even
so, he is a physicist; for he has to a high degree what makes the
physicist; the genuine interest in the experimental facts in all their 
puzzling complexity. His accurate, instructive estimate of the
relative weight of relevant experimental facts has been an unfailing
guide for him in his theoretical investigations. Pauli combines in an
exemplary way physical insight and mathematical skill.''
\end{quote}

Schliesslich soll Pauli selber nochmals zu Wort kommen. Einen Aufsatz
\"uber Einstein beschloss er mit den folgenden Worten, die genauso auf 
ihn selber passen:

\begin{quote}
``Sein in die Zukunft weisendes Leben wird uns stets gemahnen an das
in unserer Zeit bedrohte Ideal des geistigen, kontemplativen Menschen, 
dessen Gedanken ruhig und unbeirrbar den grossen Problemen der
Struktur des Kosmos nachh\"angen.''
\end{quote}


\vspace{.5cm}

\section*{Fussnoten und Referenzen}
\begin{enumerate}
\item[{[1]}] {\it 
    Collected Scientific Papers by Wolfgang Pauli.} Herausgegeben von
  R. Kronig und V.F. Weisskopf. 2 B\"ande. Wiley, New York 1964.
\item[{[2]}] P. Ehrenfest. {\it Ansprache zur Verleihung der Lorentzmedaille
  an Professor Wolfgang Pauli am 31.~Oktober
  1931.} Versl. Akad. Amsterdam {\bf 40}, 121--126 (1931).
\item[{[3]}] Wolfgang Pauli. {\it Wissenschaftlicher Briefwechsel mit
    Bohr, Einstein, Heisenberg u.a.} Band I: 1919--1929. Herausgegeben 
  von A. Hermann, K. v.~Meyenn und V.F. Weisskopf. Springer, New
  York/Heidelberg/Berlin 1979. --- Band II: 1030 bis
  1039. Herausgegeben von K. v.~Meyenn, unter Mitwirkung von
  A. Hermann und V.F. Weisskopf. Springer, Berlin/Heidelberg/New
  York/Tokyo 1985. Band IV, Teil I: 1950--1952 (Springer-Verlag,
  1996); Band IV, Teil II: 1953--1954 (Springer-Verlag 1999); Teile
  1954--1958 in Vorbereitung. Herausgegeben von K. v.~Meyenn. 
\item[{[4]}] Pauli an Einstein. Brief [239] vom 19.~Dezember 1929, in
  Ref.~[4], Band I. S.~526.
\item[{[5]}] A. Sommerfeld. {\it F\"uhrende Geister der
  naturwissenschaftlichen Forschung.} Vortrag auf dem bayrischen
  Philologentag in Regensburg. Auszugsweise abgedruckt in den
  M\"unchner Neuesten Nachrichten. September 1926.
\item[{[6]}] W. Pauli. {\it \"Uber die Energiekomponenten des
    Gravitationsfeldes}, Physikal.~Z.~{\bf 20}, 25--27 (1919); {\it Zur
    Theorie der Gravitation und der Elektrizit\"at von Hermann Weyl},
  Physikal.~Z.~{\bf 20}, 457--467 (1919); {\it Mercurperihelbewegung und
    Strahlenablenkung in Weyls Gravitationstheorie},
  Verhandl.~Deutsche Phys.~Ges.~{\bf 21}, 742--750 (1919). 
\item[{[7]}] H. Weyl. {\it Gravitation und Elektrizit\"at},
  Sitzungsber.~Deutsch.~Akad.~Wiss. Berlin, 1918, p.p.~465--480. Siehe
  auch H. Weyl, {\it Gesammelte Abhandlungen}, Hrg. K. Chandrasekharan 
  (Springer, Berlin), 1968.
\item[{[8]}] Weyl an Pauli. Brief [1] vom \dots  in Ref.~[3], Band I,
  S.~3.
\item[{[9]}] W. Pauli. {\it Relativit\"atstheorie}, in Enzyklop\"adie
  der mathematischen Wissenschaften, 5, Teil 2 (Teubner, Leipzig,
  1921), p.p.~539--775. Zum hundertsten Geburtstag von Pauli hat
  Domenico Giulini eine kommentierte Neuausgabe im Springer-Verlag
  besorgt. 
\item[{[10]}] A. Einstein. Naturwiss.~{\bf 10}, 184--185 (1922).
\item[{[11]}] W. Pauli. {\it \"Uber das Modell des
    Wasserstoffmolek\"ulions}. Ann. Physik [4], {\bf 68}, 177--240 (1922).
\item[{[12]}] A. Einstein/M. Born. {\it Briefwechsel
    1916--1955}. Kommentiert von M. Born. Reinbek 1972. Brief [35] vom 
  29.~November 1921, S.~92.
\item[{[13]}] In Ref. [13]. Kommentar zu einem Brief von Pauli an Born
  vom 11.~Dez.~1955. S.~301.
\item[{[14]}] W. Pauli. {\it \"Uber den Zusammenhang des Abschlusses der 
    Elektronengruppen im Atom mit der Komplexstruktur der
    Spektren}. Z.~Phys.~{\bf 31}, 765--783 (1925).
\item[{[15]}] Ref. [4]. Siehe speziell die Briefe [68--83].
\item[{[16]}] W. Pauli. {\it \"Uber die Gesetzm\"assigkeiten des
    anomalen Zeemaneffektes.} Z.~Physik {\bf 16}, 155--164 (1923).
\item[{[17]}] W. Pauli. {\it Exclusion principle and quantum mechanics.}
  Nobel lecture, delivered at Stockholm, December 13, 1946. (Deutsche
  \"Ubersetzung in Ref.~[1], S. 129--146.)
\item[{[18]}] Pauli an Sommerfeld. Brief [37] vom 6.~Juni 1923 in
  Ref.~[4], Band I, S.~94.
\item[{[19]}] O. Klein. {\it  Wolfgang Pauli}. Nagra Minnesord. Kosmos,
  Fysika Uppsatzer 37, 9--12 (1959). (Deutsche \"Ubersetzung in
  Ref. [7], S. 49--52.)
\item[{[20]}] W. Pauli. {\it \"Uber den Einfluss der
    Geschwindigkeitsabh\"angigkeit der Elektronenmasse auf den
    Zeemaneffekt.} Z. Physik {\bf 31}, 373--385 (1925).
\item[{[21]}] Pauli an Land\'e. Brief [71] vom 24. November 1924 in
  Ref. [4], Band I, S. 176.
\item[{[22]}] Siehe dazu B.L. van der Waerden. {\it Exclusion Principle
    and Spin.} In M. Fierz and V.F. Weisskopf: {\it Theoretical
    Physics in the Twentieth Century: A Memorial Volume to Wolfgang
    Pauli.} Interscience, New York 1960, p. 199--244. Siehe ferner
  K. v.~Meyenn: {\it Paulis Briefe als Wegbereiter wissenschaftlicher
    Ideen}. In Ref. [7], S. 20--42.
\item[{[23]}] G.E. Uhlenbeck und S. Goudsmit. {\it Ersetzung der
    Hypothese vom unmechanischen Zwang durch eine Forderung
    bez\"uglich des inneren Verhaltens jedes einzelnen Elektrons.}
  Naturwiss. 13, 953 (1925). {\it Spinning Electrons and the Structure
    of Spectra.} Nature, Lond. {\bf 117}, 264 (written December 1925).
\item[{[24]}] E.C. Stoner. {\it The Distribution of Electrons among
    Atomic Levels.} Philosophical Magazine {\bf 48}, 719--736 (1924).
\item[{[25]}] Pauli an Bohr. Brief [74] vom 12. Dezember 1924 in
  Ref. [4], Band I, S. 186--189.
\item[{[26]}] W. Pauli. {\it Zur Frage der theoretischen Deutung der
    Satelliten einiger Spektrallinien und ihrer Beeinflussung durch
    magnetische Felder.} Naturwiss. {\bf 12}, 741--743 (1924).
\item[{[27]}] W. Pauli. {\it Zur Quantenmechanik des magnetischen
    Elektrons.} Z. Physik {\bf 43}, 601--623 (1927).
\item[{[28]}] Heisenberg an Pauli. Brief [103] vom 3. November 1925 in
  Ref. [4], Band I, S. 252.
\item[{[29]}] W. Heisenberg. {\it Erinnerungen an die Zeit der
    Entwicklung der Quantenmechanik.} In M. Fierz and V.F. Weisskopf:
  Theoretical Physics in the Twentieth Century: A Memorial Volume to
  Wolfgang Pauli. Interscience New York 1960, p. 40--47. 
\item[{[30]}] Pauli an Heisenberg. Brief [143] vom 19. Oktober 1926 in
  Ref. [4], Band I, S. 340.
\item[{[31]}] Pauli an Jordan. Brief [131] vom 12. April 1926 in
  ref. [4], Band I, S. 315.
\item[{[32]}] R. Jost. {\it Buchbesprechung ``Wolfgang Pauli,
    Wissenschaftlicher Briefwechsel''}, Band I: 1919--1929. ZAMP {\bf 31},
  548 (1980).
\item[{[33]}] P. Jordan und W. Pauli. {\it Zur Quantenelektrodynamik
    ladungsfreier Felder.} Z. Physik {\bf 47}, 151--173 (1928).
\item[{[34]}] W. Heisenberg und W. Pauli. {\it Zur Quantenelektrodynamik 
    der Wellenfelder.} Z. Physik {\bf 56}, 1--61 (1929) und W. Heisenberg
  und W. Pauli. {\it Zur Quantentheorie der Wellenfelder~II.}
  Z. Physik {\bf 59}, 168--190 (1930).
\item[{[35]}] W. Pauli. {\it Die allgemeinen Prinzipien der
    Wellenmechanik.} Handbuch der Physik, 2. Auflage, Bd. 24,
  1. Teil. S. 83--272. Springer, Berlin 1933. 
\item[{[36]}] W. Pauli und V. Weisskopf. {\it \"Uber die Quantisierung
    der skalaren relativistischen Wellengleichung.} Helv. Phys. Acta
  {\bf 7}, 709--731 (1934). 
\item[{[37]}] Pauli an Heisenberg. Brief [375] vom 28. Juni 1934, in
  Ref. [4], Band II, S. 334.
\item[{[38]}] Pauli an Heisenberg. Brief [394] vom 7. November 1934, in
  Ref. [4], Band II, S. 360.
\item[{[39]}] P.A.M. Dirac. Proc. Rov. Soc. A {\bf 155}, 447 (1936).
\item[{[40]}] M. Fierz. Helv. Phys. Acta {\bf 12}, 3 (1939); {\bf 23}, 412 (1950).
\item[{[41]}] W. Pauli. {\it The connection between spin and
    statistics.} Phys. Rev. {\bf 58}, 716--722 (1940).
\item[{[42]}] R. Kronig. Naturw. {\bf 16}, 335 (1928).
\item[{[43]}] W. Heitler und G. Herzberg. Naturw. {\bf 17}, 673 (1929).
\item[{[44]}] L. Meitner und W. Orthmann. Z. Physik, {\bf 60}, 143 (1930).
\item[{[45]}] W. Pauli. {\it Offener Brief an die Gruppe der Radiaktiven 
    bei der Gauvereins-Tagung zu T\"ubingen, 4. Dezember 1930.}
  (Wiedergegeben in Ref. [1], S. 159.)
\item[{[46]}] Pauli an Klein. Brief [261] vom 12. Dezember 1930, in
  Ref. [4], Band II, S. 43--47.
\item[{[47]}] W. Pauli. {\it Exclusion principle, Lorentz group and
    reflection of space-time and charge. Niels Bohr and the
    Development of Physics.} W. Pauli, Hrsg. Pergamon, London 1955,
  S.~30--51.
\item[{[48]}] R. Jost. Helv. Phys. Acta {\bf 30}, 409 (1957). Siehe auch
  R. Jost. {\it Das Pauli-Prinzip und die Lorentz-Gruppe.} In M. Fierz 
  and V.F. Weisskopf. Theoretical Physics in the Twentieth
    Century: A Memorial Volume to Wolfgang Pauli. Interscience New
  York 1960, p. 107--136. 
\item[{[49]}] W. Pauli and F. Villars. {\it On the invariant
    regularization in relativistic quantum theory.}
  Rev. Mod. Phys. {\bf 21}, 434--444 (1949).
\item[{[50]}] Pauli an Pais. Brief [1614] und [1682] in Ref. [3], Band
  IV, Teil 2.
\item[{[51]}] Pauli an Yang. Brief [1726] vom 24. Feb. 1954, Band IV,
  Teil 2.
\item[{[52]}] P. Gulmanelli. {\it Su una Teoria dello Spin Isotopico,}
  Pubblicazioni della Sezione di Milano dell' Instituto Nazionale di
  Fisica Nucleare, Casa Editrice Pleion, Milano (1954).
\item[{[53]}] L. O'Raifeartaigh, N. Straumann. Reviews of Modern
  Physics, {\bf 72}, 1--23 (2000).
\item[{[54]}] H. Weyl (1946): Memorabilia,  in {\it Hermann Weyl,}
  Hrg. K. Chandrasekharan (Springer, New York, 1986), p. 85.
\item[{[55]}] W. Pauli. {\it Relativit\"atstheorie und Wissenschaft,} in 
  Ref. [56], S. 76--80.

\item[{[56]}] W. Pauli. {\it Aufs\"atze und Vortr\"age \"uber Physik und
Erkenntnistheorie}. Vieweg, Braunschweig 1961. Die Trauerrede von
Weisskopf er\"offnet diesen Band. Eine Neuauflage mit einleitenden
Bemerkungen von K. v.~Meyenn erschien als Band 15 der Facetten der
Physik, hrsg. von R.U. Sexl, unter dem Titel {\it Physik und
Erkenntnistheorie}. Vieweg, Braunschweig 1984.
\item[{[57]}] Wolfgang Pauli. {\it Das Gewissen der Physik}. Hrsg.:
  Ch.~P. Enz, K. v.~Meyenn, Vieweg, Braunschweig 1988. Der Brief von
  Pauli ist in der Einleitung von Ch.~Enz abgedruckt. Dieser Band
  enth\"alt neben pers\"onlichen Erinnerungen und Ankedoten eine
  Auswahl von Paulis wichtigsten Arbeiten. In wertvollen Anh\"angen
  findet man eine Zeittafel, ein Verzeichnis der Schriften von
  Wolfgang Pauli sowie der bestehenden Sekund\"arliteratur.
\end{enumerate}

\vspace{1cm}

\end{document}